\documentclass[a4paper,fleqn,usenatbib]{mnras}

\usepackage{newtxtext,newtxmath}
\usepackage[T1]{fontenc}
\usepackage{ae,aecompl}

\usepackage{graphicx}	% Including figure files
\usepackage{amsmath}	% Advanced maths commands
\title[HORuS transmission spectroscopy of 55~Cnc~e]{HORuS transmission spectroscopy of 55~Cnc~e}

\author[H. M. Tabernero et al.]{
H. M. Tabernero,$^{1,2}$\thanks{E-mail: hugo.tabernero@astro.up.pt}%
C. Allende Prieto,$^{3,4}$
M. R. Zapatero Osorio,$^{2}$
\newauthor
J. I. Gonz{\'a}lez Hern{\'a}ndez,$^{3,4}$
C. del Burgo,$^{5,4}$
R. Garc{\'i}a L{\'o}pez,$^{3,4}$
R. Rebolo, $^{3,4}$
\newauthor
M. Abril-Abril,$^{6}$
R. Barreto,$^{3,4}$
J. Calvo Tovar,$^{3,4}$
 A. D{\'i}az Torres,$^{3,4}$
\newauthor
P. Fern{\'a}ndez Izquierdo,$^{3,4}$
M.F. G{\'o}mez-Re\~{n}asco,$^{3,4}$
F. Gracia-T{\'e}mich,$^{3,4}$
E. Joven,$^{3,4}$
\newauthor
J. Pe\~{n}ate Castro,$^{3,4}$   
S. Santana-Tschudi,$^{3,4}$  
F. Tenegi,$^{3,4}$  
and H. D. Viera Mart{\'i}n$^{6}$  
\\
$^{1}$Instituto de Astrof{\'i}sica e Ci{\^e}ncias do Espa\c{c}o, Universidade do Porto, CAUP, Rua das Estrelas, 4150-762 Porto, Portugal \\
$^{2}$Centro de Astrobiolog{\'i}a (CSIC-INTA), Carretera de Ajalvir km 4, Torrej{\'o}n de Ardoz, 28850 Madrid, Spain \\
$^{3}$Dpto. Astrof\'{i}sica, Universidad de La Laguna, 38206 La Laguna, Tenerife, Spain \\
$^{4}$Instituto de Astrof\'{i}sica Canarias, v\'{i}a L\'{a}ctea s/n, 38205 La Laguna, Tenerife, Spain\\
$^{5}$Instituto Nacional de Astrof\'{i}sica, ́ \'{O}ptica y Electr\'{o}nica, Luis Enrique Erro 1, Sta. Ma. Tonantzintla, Puebla, Mexico\\
$^{6}$Gran Telescopio Canarias,  Cuesta de San Jos{\'e}, s/n, 38712, Breña Baja, La Palma, Spain 
}

\date{Accepted: 2020 August 19; Received: 2020 January 6}

\pubyear{2020}

\begin{document}
\label{firstpage}
\pagerange{\pageref{firstpage}--\pageref{lastpage}}
\maketitle

\begin{abstract}
The High Optical Resolution Spectrograph (HORuS) is a new high-resolution {\it echelle} spectrograph available on the 10.4~m  Gran Telescopio Canarias (GTC). We report on the first HORuS observations of a transit of the super-Earth planet 55~Cnc~e.  We investigate the presence of \ion{Na}{i} and H$\alpha$ in its transmission spectrum and explore the capabilities of HORuS for planetary transmission spectroscopy. Our methodology leads to residuals in the difference spectrum between the in-transit and out-of-transit spectra for the \ion{Na}{I} doublet lines of (3.4~$\pm$~0.4)~$\times$~10$^{-4}$, which sets an upper limit to the detection of line  absorption from the planetary atmosphere that is one order of magnitude more stringent that those reported in the literature. We demonstrate that we are able to reach the photon-noise limit in the residual spectra using HORuS to a degree that we would be able to easily detect giant planets with larger atmospheres.  In addition, we modelled the structure, chemistry and transmission spectrum of 55~Cnc~e using state-of-the-art open source tools.

\end{abstract}


%%%%%%%%%%%%%%%%%%%%%%%%%%%%%%%%%%%%%%%%%%%%%%%%%%

%%%%%%%%%%%%%%%%% BODY OF PAPER %%%%%%%%%%%%%%%%%%

\section{Introduction}

Planets orbiting stars are ubiquitous, as demonstrated by more  than 4000 planets discovered to date\footnote{\url{https://exoplanetarchive.ipac.caltech.edu}}. With the pace of exoplanetary discoveries still increasing thanks to multiple transiting and radial velocity surveys carried out with e.g. space missions like {\it Kepler, K2, CoRoT}, and {\it TESS} \citep{ric14}; the ground-based spectrographs such as HARPS \citep{may03}, HARPS-N \citep{cos12},  HIRES \citep{vog94}, CARMENES \citep{quir16}, and MARVELS \citep{ala15}; and new and future instruments like ESPRESSO \citep{pepe10}, CHEOPS \citep{ran18},  HPF \citep{hpf}, JWST \citep{gar06},  PLATO \citep{rau14}, NIRPS \citep{wil17}, and SPIRou \citep{mou15}. Exoplanetary research is now approaching the deep study and characterization of the atmospheres of the extrasolar planets.\\

Transiting exoplanets present a unique opportunity to characterize their atmospheres, and learn about their chemical composition. \citet{char02} reported the first detection of an exoplanet atmosphere by means of Hubble Space Telescope spectroscopy. These authors detected the resonance doublet of \ion{Na}{i} in the Jupiter-size planet HD~209458~b.  \citet{cas20} showed that this detection is due to the Rossiter-McLaughlin (RM) effect and not to the atmosphere of the planet. In spite of that result, \citet{char02} opened a new era in the study of exoplanets. The first detection of an exoplanet atmosphere with a ground-based telescope was made by \citet{red08}, who identified the absorption of the sodium resonance lines caused by the atmosphere of HD~189733~b. Line absorption of Na, K, He, CO, CH$_{\rm 4}$, or water vapour has  already been reported in the atmospheres of tens of giant gaseous planets \citep[e.g.,][]{bar15,all17,she17,chen18,nor18,par18,alo19,sei19}.   In addition, atomic absorption features due to \ion{Fe}{i,ii} and \ion{Ti}{i,ii} have been detected in the atmosphere of the ultra-hot Jupiter KELT-9~b \citep{hoe18,hoe19}. These measurements represent a step forward in our understanding of the chemistry of planetary atmospheres and our ability to constrain planet formation theories. The study of exoplanetary  atmospheres is also moving towards the exploration of the atmospheres of rocky planets (i.e., super-Earths), but these atmospheres are harder  to detect. \\

Spectroscopy is a powerful tool used to detect both broad and narrow features in the  transmission spectra of transiting exoplanets, all of which are important ingredients to unveil their chemistry. The High Optical Resolution Spectrograph (HORuS) at the Gran Telescopio Canarias (GTC) offers an excellent opportunity for this task, given its moderate spectral resolution (R~$\approx$~25,000) at optical wavelengths, and large collecting area, which is critical for achieving a high signal-to-noise ratio (S/N) during the limited time window of the planetary transit.\\

In this work, we analyse the HORuS commissioning data for a single transit of the hot and dense super-Earth planet 55~Cnc~e, which is orbiting one of the brightest planet-host stars known to have a super-Earth planet \citep{bou18}. The 55~Cnc system is an interesting multiplanetary system with five known planets that have been discovered during the last decades \citep[see][]{55cncb,55cncce,55cncd,55cncf}. 55~Cnc~e is a supereath planet of 7.99~M$_{\oplus}$ and a planetary radius of 1.875~$R_{\oplus}$. It orbits its parents star with a short period of $\approx$~0.7~d at a distance to the host star 55~Cnc of 0.01~au. This paper consists of the following sections: The  HORuS observations and data reduction procedure  are described in Sect.~\ref{sect_obs}. The extraction of the planetary signal is documented in Sect.~\ref{sect_analysis}. In Sect.~\ref{sect_dis} we discuss our results and compare them to those obtained in previous works. Finally, in Sect.~\ref{sect_con} we present the impact of our findings.\\

\section{Observations} 

\label{sect_obs}

A series of spectroscopic observations covering one transit of the super-Earth 55~Cnc~e  (see Table~\ref{par_planet}) was obtained with HORuS\footnote{\url{http://research.iac.es/proyecto/abundancias/horus/index.php}}, which is a moderate resolution {\it echelle} spectrograph (R~$\approx$~25,000) in operation at the 10.4~m Gran Telescopio Canarias (GTC). It collects light at the Nasmyth-B focal plane, shared with OSIRIS, using a 3$\times$3 integral field unit (IFU, 2.1$\times$2.1 arcsec) with microlenses, into optical fibres that form a pseudo-slit at the spectrograph entrance, fitted with another set of microlenses. The light is dispersed with a 79 gr~mm$^{-1}$ {\tt echelle} grating and cross-dispersed with three prisms, providing nearly continuous coverage between 3800 and 6900~\AA~split over 27 spectral orders. The detector is a 4096$\times$4096 15$-\mu m$-pixel Fairchild CCD.\\

We observed one transit of 55~Cnc~e using HORuS at the GTC during the night of 12 Dec 2018. This night and the preceding one were part of a commissioning run for HORuS. The CCD was read with no binning, and a readout time of 90~s. With a sampling of 6 pixels per resolution element in the spectral direction, and about the same number of pixels sampling each of the nine fibres in the IFU in the spatial direction, the 16-bit electronics can provide a maximum signal-to-noise ratio of several thousand per exposure. We took spectra of the Th-Ar hollow-cathode for wavelength calibration during the transit, with a cadence of approximately 1~h. We obtained four 900~s exposures followed by twenty 450~s integrations, spanning slightly more than 4~h, and including the transit. Each 450~s exposure provided a signal-to-noise per resolution element in excess of 1800 at 550 nm, and about half of that at 420 nm. The observations span over $\approx$~4.2~h, whereas the transit lasted for $\approx$~1.6~h.  A total of 10 spectra were acquired during the planetary transit of 55~Cnc~e. In addition, we started observing at an airmass of 1.55, whereas the observations finished at an airmass of 1.05. \\
\begin{table}
%\centering
\caption{Orbital and physical parameters for 55~Cnc and 55~Cnc~e}            % title of Table
\label{par_planet}      
\begin{tabular}{l c c }   
\hline
\hline                 
Parameter &  Value & Reference \\    
\hline                       
\hline
& Stellar properties & \\
 \hline
   $T_{\rm eff}$ & 5353~$\pm$~62~K & \citet{sou18} \\      
   $\log{g}$ &  4.30~$\pm$~0.14~dex&   \citet{sou18} \\
   $\lbrack$Fe/H$\rbrack$  & 0.30~$\pm$~0.04~dex   &  \citet{sou18} \\ % AFesun = 7.50 dex 
   $\xi$  &  1.01~$\pm$~0.10~km~s$^{-1}$ & \citet{sou18} \\
   $M_{*}$   &  0.97~$\pm$~0.09~$M_{\odot}$  &  \citet{bou18} \\
   $R_{*}$   &  1.73~$\pm$~0.04~$R_{\odot}$ &   \citet{bou18} \\
\hline
& Planet properties &\\
\hline
   $P$        & 0.7365474~$\pm$~0.0000014 d & \citet{bou18}\\ 
   $M_{p}$  &  7.99~$\pm$~0.33~$M_{\oplus}$  & \citet{bou18} \\
   $R_{p}$  &  1.875~$\pm$~0.05~$R_{\oplus}$  &  \citet{bou18}  \\
   $K_1$    &  6.02~$\pm$~0.24~m~s$^{-1}$ &  \citet{bou18} \\
   $a$        &  0.01544~$\pm$~0.00005~au & \citet{bou18}\\  
   $\gamma$ &  27.45145~$\pm$~0.00089~km~s$^{-1}$ & \citet{bou18}\\
   $e$      & 0.05~$\pm$~0.03  & \citet{bou18}  \\
   $\omega$  & 86.0$^{\circ}$~$\substack{+30.7 \\ -33.4}$ &  \citet{bou18} \\
\hline
\end{tabular}
\end{table}

\section{Data analysis}
\label{sect_analysis}
\subsection{Data reduction}

The {\tt chain}\footnote{\url{https://github.com/callendeprieto/chain}}, a  custom-made data reduction pipeline for HORuS written in IDL, was used to process the data, performing bias removal,  finding, tracing, and extraction of the apertures corresponding to the spectral orders. Wavelength calibration involved fitting third or -order polynomials to one to two dozen thorium lines in each {\it echelle} order, identified in an atlas made with the Tull spectrograph \citep{cap01}, and interpolating linearly the wavelength solutions between the Th-Ar exposures preceding and following each stellar spectrum. The collapse of the orders along the slit was done by adding the counts along the spatial direction of the detector. By taking ratios of two consecutive out-of-transit spectra we find that the distribution of the residuals for the central regions of the orders in the red part of the spectrum is well described by a Gaussian with a $\sigma \simeq 0.0018$, which is equivalent to a signal-to-noise ratio per pixel of $\sqrt{2}/0.0018 \simeq 800$, or 1900 per resolution element, confirming the expectations based on the Poisson noise ($\sqrt{N_{counts}}$) for the accumulated stellar signal.

\subsection{Instrumental corrections}

Before attempting the extraction of the absorption signature imprinted by the planet's atmosphere in the spectrum of the star, we decided to analyse two particular spectral regions: the \ion{Na}{i} doublet (5889.95 and 5895.92~\AA) and the H$\alpha$ line (6562.80~\AA). {HORuS is not stabilized thermally or mechanically}, as it is the case of, e.g., ESPRESSO \citep{pepe10}, HARPS \citep{may03}, and CARMENES \citep{quir16}. Consequently, variations in the point spread function of the instrument, inducing changes of up to 10\% in scales of hours, are typically found. We found that stability issues can be addressed by carefully aligning all the observed spectra by working in small wavelength regions of interest. Each individual spectroscopic observation has to be aligned with the others in terms of wave and flux. Spectral alignment is critical to extract the planetary signature, as the wavelength solution must be as precise as possible. To that aim, we generated an \ion{Fe}{i} synthetic spectrum by means of {\tt turbospectrum}\footnote{\url{https://github.com/bertrandplez/Turbospectrum2019}} \citep{ple12}. In addition, we used a {\tt MARCS} model  atmosphere \citep{gus08} of T$_{\rm eff}$~$=$~5250~K, $\log{g}$~$=$~4.5~ dex, and [M/H]~$=$~0.25~dex\footnote{The closest {\t MARCS} model to the parameters given in Table~\ref{par_planet}}, alongside an \ion{Fe}{i} line list extracted from the Vienna Atomic Line Database \citep[VALD3;][]{rad15}. We adopted {\tt iSpec} \citep{bla14} to cross-correlate each individual stellar observation with the generated \ion{Fe}{i} synthetic spectrum. As a result we could measure the radial velocity corrections ($\sigma$~$\approx$~0.2~km~s$^{-1}$, see Table~\ref{tab:obs_rv_table}) required to shift each individual spectrum to the rest frame. \\

Variations in the effective spectral resolution of HORuS also have to be taken into account to extract the planetary signal. Thus, we calculated a nominal resolution by measuring the width of Th lines in the calibration frames.  The full width at half-maximum ($FWHM$) of the cross-correlation given by {\tt iSpec} of each individual Th-Ar exposure with itself (divided by $\sqrt{2}$) will give us an effective resolution for each exposure. At this point, we can homogenize all the observations to the exact same resolution, by means of convolving each spectrum with a Gaussian kernel that accounts for the missing broadening.\\

\subsection{Extraction of the planetary signal}

\begin{figure}                                                                                                                                                                                            \includegraphics[width=\columnwidth]{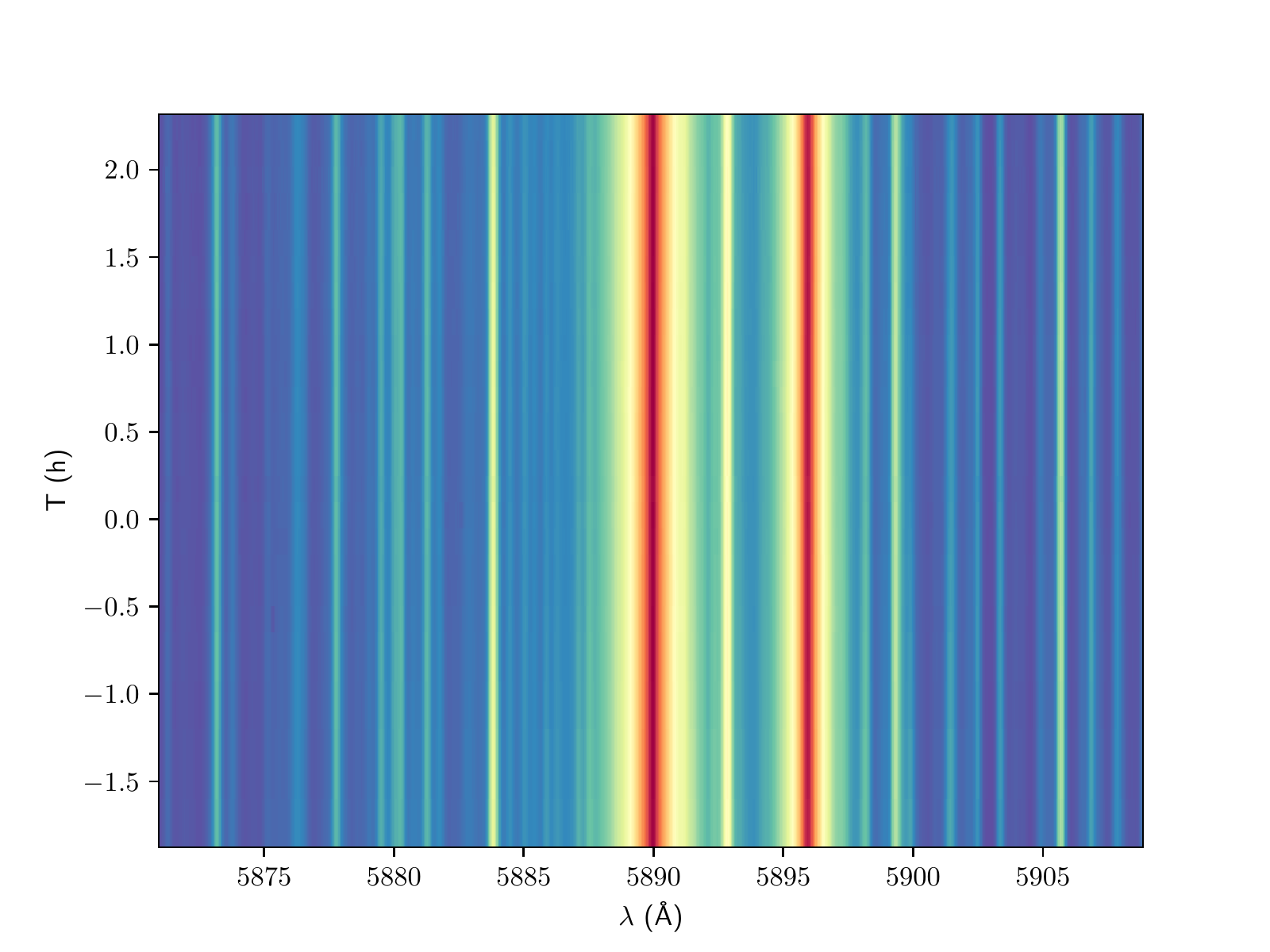}
\caption{All HORuS spectra are shown vertically ordered in time (the earliest spectra are at the bottom). Time 0.0 h corresponds to the mid-transit according to the planetary ephemeris given in Table~\ref{par_planet}. The blue colour represents the continuum and all other colours stand for atomic absorption features (\ion{Na}{i} doublet lines are the  deepest). All stellar features are well aligned vertically and the flux continuum is nearly identical for all spectra.}                                                                   
\label{fig:scale}                                                                                                                                                                              
\end{figure}                                

The flux changes from exposure to exposure, due to variations in airmass and atmospheric transparency. Individual exposures must be corrected for these effects, and we used a  low-order polynomial in order to have set of spectra in the same flux scale. The final result of flux alignment and wavelength correction is shown in Fig.~\ref{fig:scale}. We separated the spectra into two categories, in-transit and out-of-transit spectra, by means of the ephemeris given by \citet{bou18}. Then, we computed,  wavelength by wavelength, the median of the out-of-transit spectra to generate a master spectrum.  We employed the median to compute the out-of-transit spectrum because it is a non-skewed estimator more resilient to the influence of "bad pixels" (i.e. cosmic rays) in an individual exposure. We can divide each observation by this master spectrum to get a set of residual spectra that can be later combined to extract the signature of the planetary transit.\\  

The telluric correction was done in an iterative manner, since the wavelength ranges of interest to this work contain heavily blended telluric lines to correct for (as well as the instrument resolution). However, it is possible to correct them without completely removing the planetary signature. Thus, we decided to employ the {\tt sysrem} algorithm  \citep{tam05,maz07} to correct these lines.  {\tt sysrem} is an iterative detrending algorithm that is used to remove artefacts in any set of residual spectra (after diving by the master spectrum).  This algorithm has been  successfully employed in the literature \citep[e.g.,][]{alo19,haw18,nug17} to retrieve planetary signatures. We needed five {\tt sysrem} iterations to remove the remaining telluric residuals and achieving the photon-noise limit. This gives us a set of residual spectra clean of any tentative remaining stellar and telluric residuals. We can combine the in-transit spectra after correcting them for the velocity of the planet derived using the ephemeris given by \citet{bou18} (see Table~\ref{par_planet}). After all the residual in-transit spectra are shifted to the planet rest frame we can merge them  (using a median, in a wavelength by wavelength basis) to extract the final planetary signal. \\

\begin{figure*}
\includegraphics[width=\columnwidth]{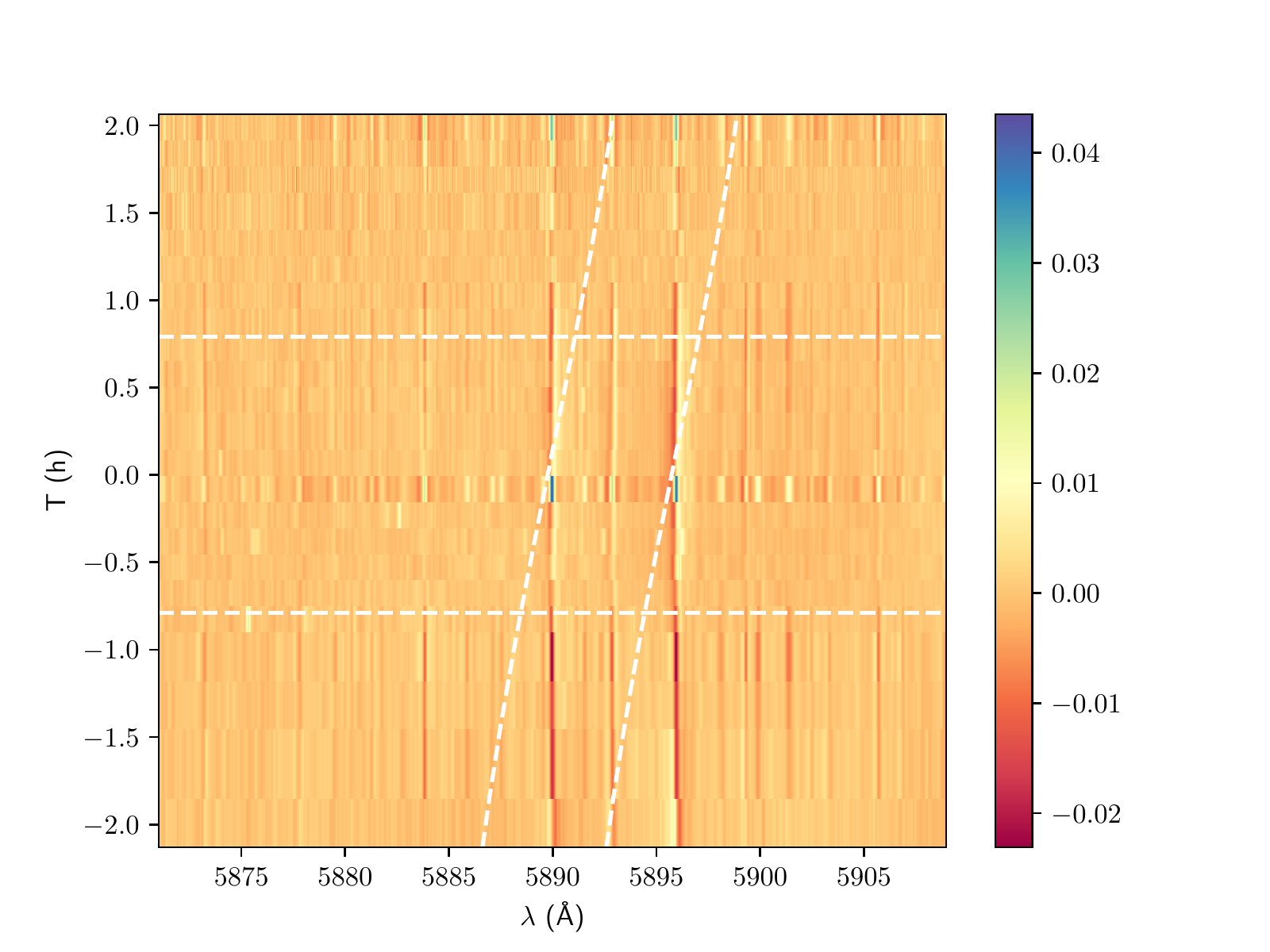}
\includegraphics[width=\columnwidth]{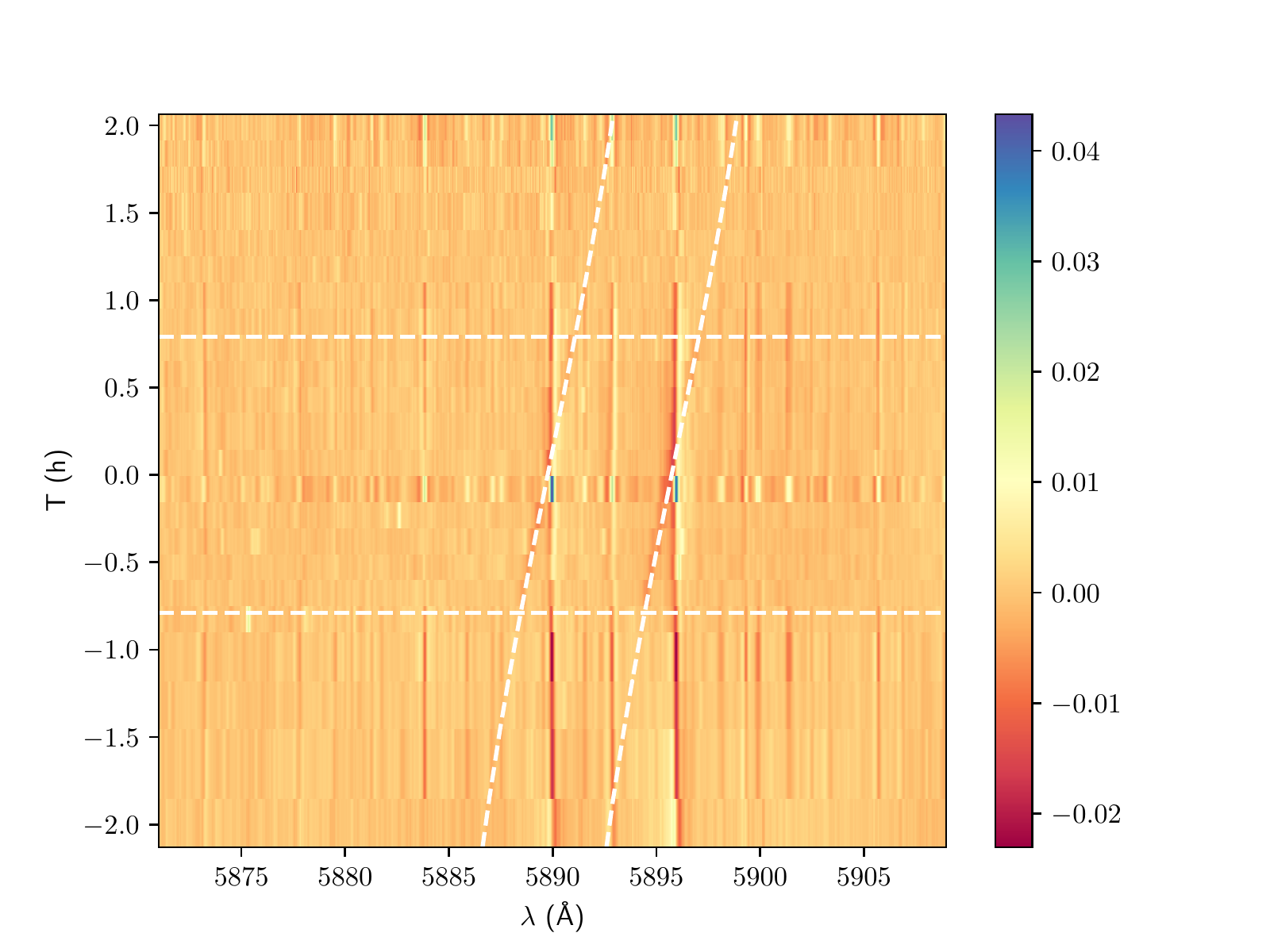}
\includegraphics[width=\columnwidth]{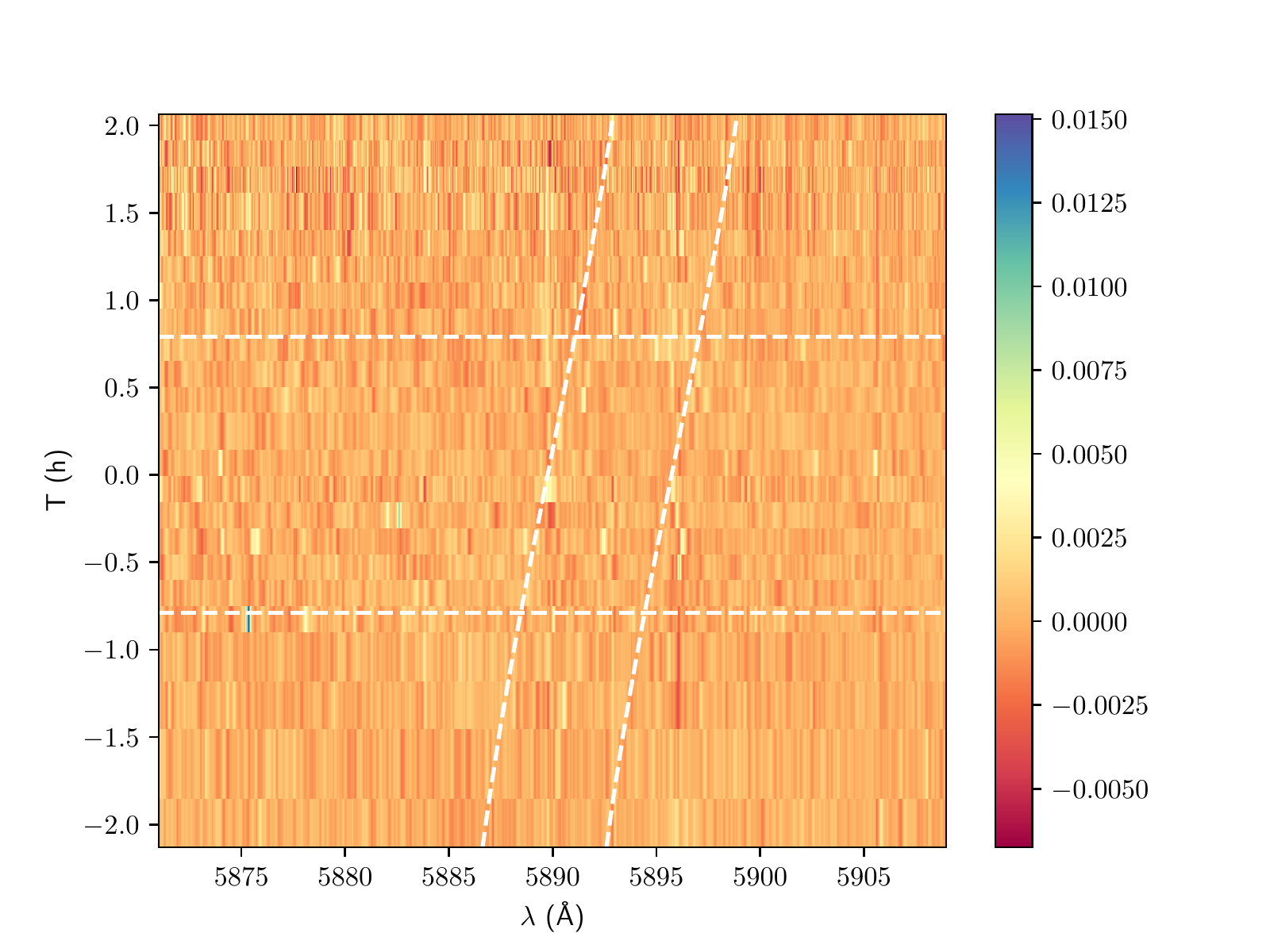}                                                                                                                                
\includegraphics[width=\columnwidth]{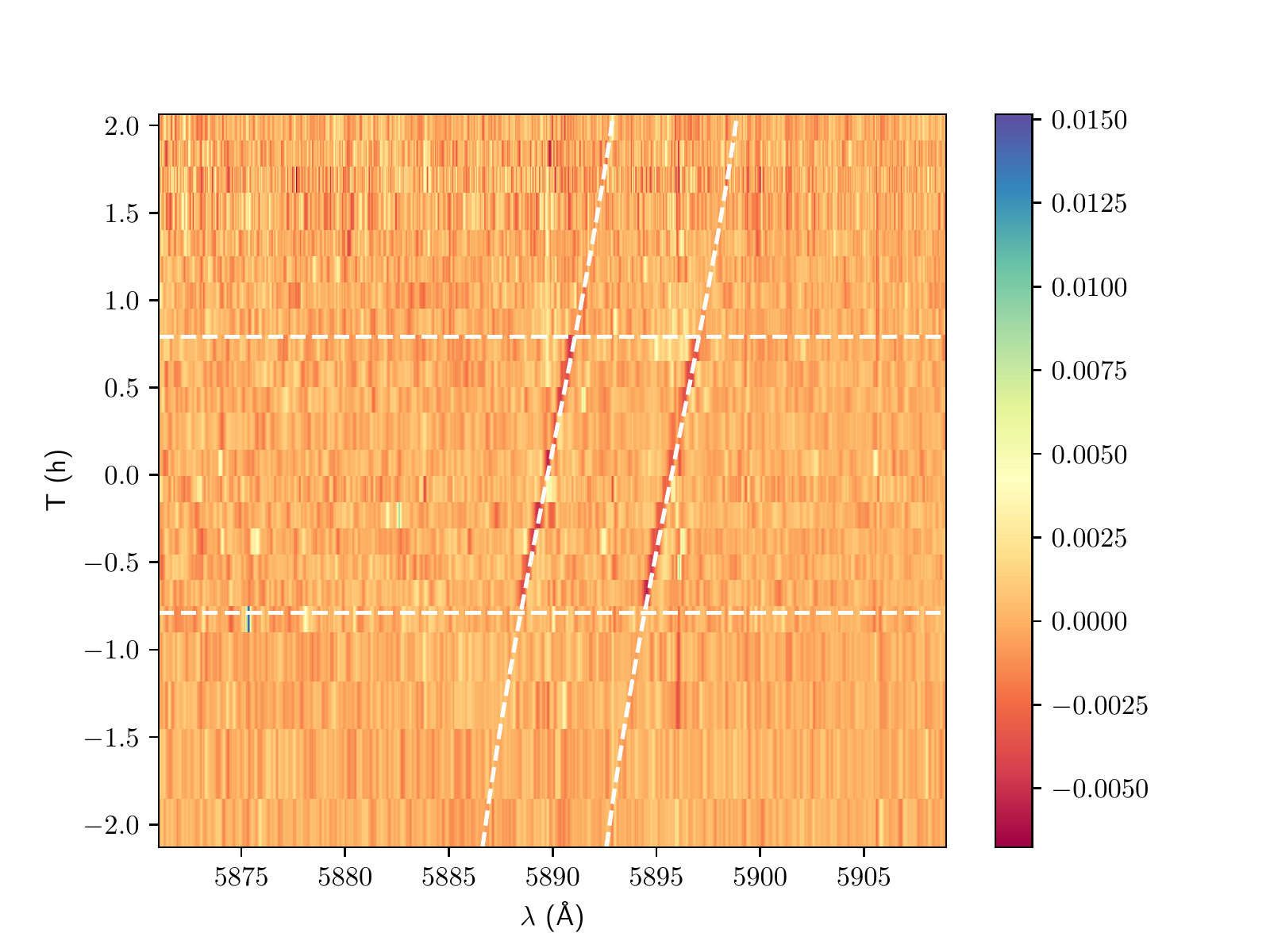}
\includegraphics[width=\columnwidth]{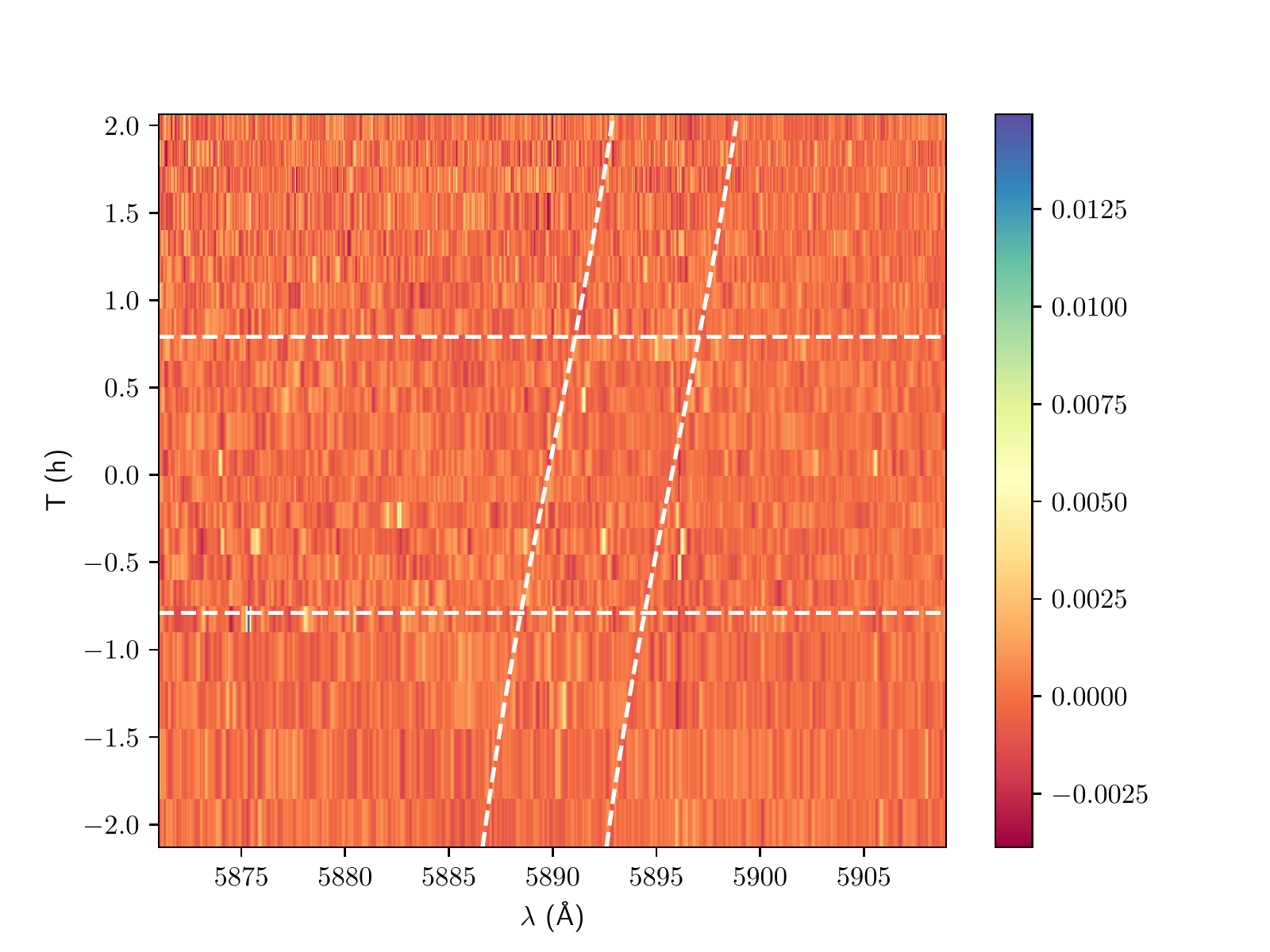}
\includegraphics[width=\columnwidth]{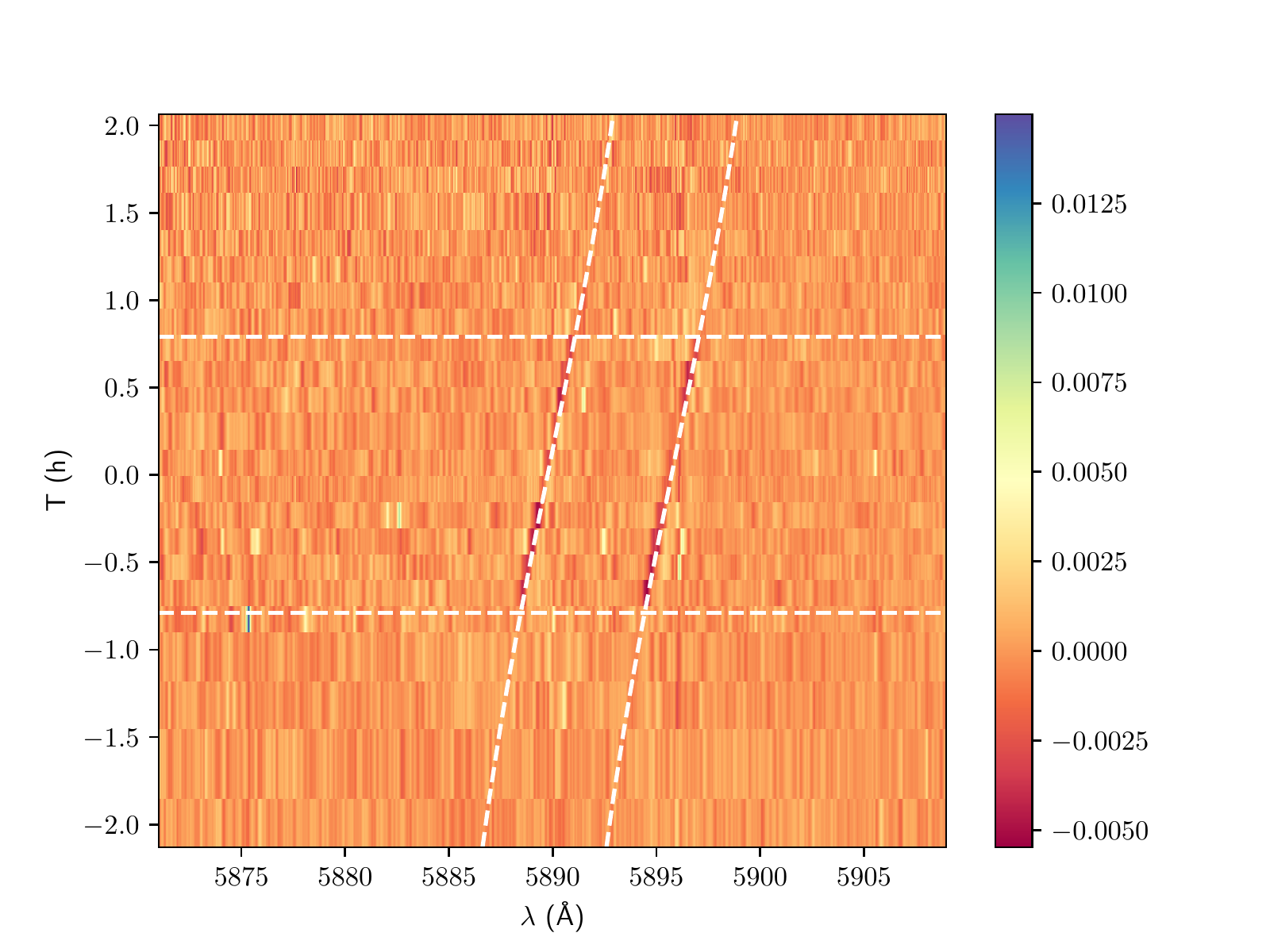}
 
\caption{Visual representation of our {\tt sysrem} implementation applied to the \ion{Na}{i} doublet. The left-hand column represents the HORuS data, whereas the right-hand column shows the same data with an injected planetary signal. Each row represents different steps of our {\tt sysrem} implementation. Here we represent our results for the removal of the master spectrum (top panels), as well as three and five {\tt sysrem} iterations. The transit window is given by the two horizontal dashed lines, whereas the planet trace is drawn by the two vertical curves.}                                                                  
\label{fig:sysrem}
\end{figure*}

It is important to mention that {\tt sysrem} can partially remove the planet signal during each iteration. In consequence, we need to check that our method is  not completely removing the planetary signal. Following \citet{ridh16} we injected a simulated planetary signal with an intensity of 0.005 to mimic the \ion{Na}{i} lines. In Fig.~\ref{fig:sysrem} we show how our method does not remove any planetary signal, and that we can recover the injected model by means of only five {\tt sysrem} iterations. We performed the analysis described above for both \ion{Na}{i} doublet and the H$\alpha$ line, we retrieve an absorption depth at about (4.1~$\pm$~0.3)~$\times$~10$^{-3}$ for both \ion{Na}{i} lines. Whereas for H$\alpha$ we retrieve a similar value of (4.4~$\pm$~0.7)~$\times$~10$^{-3}$. In consequence, these two retrieved values reveal that {\tt sysrem} leaves the majority of the artificially injected planet signature intact. 

\begin{figure}
  \includegraphics[width=\columnwidth]{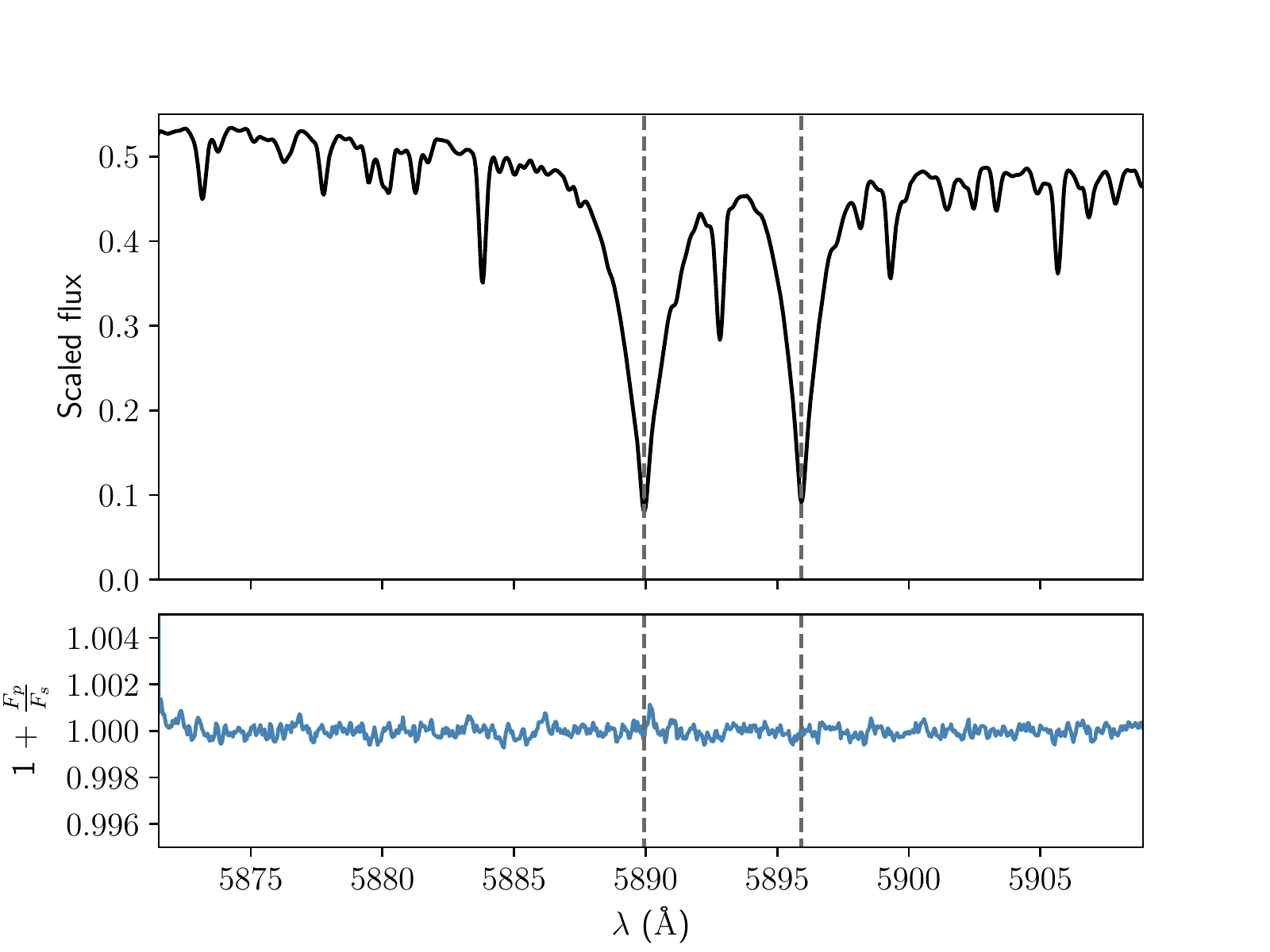}                                                                                                                               
  \caption{Final master stellar spectrum (top) and final transmission spectrum (bottom) for 55~Cnc~e in the \ion{Na}{i} region. The sodium doublet wavelengths are marked by the two vertical dashed lines.}
  \label{fig:final_spec}
  \end{figure}

\section{Discussion}

\label{sect_dis}

The resulting planetary transmission spectrum centred at the resonance doublet of \ion{Na}{i} and H$\alpha$ is shown in Figs.~\ref{fig:final_spec} and \ref{fig:final_spec_Ha}. We have not detected  \ion{Na}{i} in 55~Cnc~e, as expected from its  high mean molecular weight atmosphere contributing to only a small per cent of its planetary radius  \citep{bou18}. However, we can infer an upper limit  to the presence of these two  lines. We  have performed a bootstrap process to calculate the upper limit to the presence of the \ion{Na}{i} and H$\alpha$ lines. To that aim, we made groups of seven spectra to generate 120  transmission spectra in both spectral regions. Unfortunately, we have only a few observations and we had to make a compromise performing simulated spectra. Moreover, we need to allow variation to perform around one hundred simulations without affecting the statistics of each individual simulation. All in all,  thanks to these simulations we can calculate the accuracy  achievable using HORuS data.\\

Our calculations gave us an upper limit of (3.4~$\pm$~0.4)~$\times$~10$^{-4}$ to the presence of the \ion{Na}{i} doublet lines. This value represents a factor of $\approx$~10 improvement with respect to the previous measurements  reported by \citet{ridh16}. In addition, we have inferred  an upper limit of (7~$\pm$~1)~$\times$~10$^{-4}$ for the H$\alpha$ line. The inferred upper limits are accurate to the first significant digit according to our bootstrap calculations. This is due to the relatively small number of statistically independent samples that can be drawn by means of the bootstrap method. The previous measurements in the \ion{Na}{i} wavelength range were based on four transits of 55~Cnc~e observed with HARPS/ESO 3.6 m and HARPS-N/TNG. Despite the non-detection of Na in 55~Cnc~e, we have demonstrated that we are able to  achieve the photon-noise limit of the exoplanetary data using HORuS to a degree that we will be able to detect giant planets. These planets have larger atmospheres and transmission depths above 10$^{-3}$ (at the 3~$\sigma$ level) that can be easily detected, as indicated by our results. \\

In addition, our upper limit on H$\alpha$ is in good agreement with the conclusions  of \citet{ehr12}, as they do not detect any hints of neutral hydrogen in the transmission spectrum of 55~Cnc~e. However, \citet{tsi16} detected two tentative HCN features in the infrared wavelength range (at 1.42 and 1.54 $\mu$m) suggesting a lightweight atmosphere.\\

Our estimated upper limit to the depth of the \ion{Na}{i} doublet lines and H$\alpha$ needs to be compared to the predictions of theoretical models. To that aim we computed an atmospheric structure using the HELIOS\footnote{\url{https://github.com/exoclime/HELIOS}} code \citep[see Figs.~\ref{fig:model_atm_pt} and \ref{fig:model_atm_pz}]{mal17,mal19}. HELIOS allowed us to generate tailored models taking into account the irradiation of the parent star. HELIOS requires fundamental parameters for both the star and the planet, as well as a chemical composition. Estimates for those parameters were taken from \cite{bou18}, and we assumed  solar abundances from \citep{asp09}. The HELIOS model was used in FastChem\footnote{\url{https://github.com/exoclime/FastChem}} \citep{sto18}  to compute the number densities for several atomic and molecular species in the atmosphere of the planet (see Fig.~\ref{fig:model_atm_chem}). In addition, we calculated the transmission spectrum of 55~Cnc~e thanks to the petitRADTRANS\footnote{\url{https://petitradtrans.readthedocs.io/en/latest/}} code \citep{mol19} alongside the HELIOS model and the particle number densities we have calculated with FastChem. We used the high-resolution option of petitRADTRANS to compute two synthetic spectra of 55~Cnc~e using two different C/O ratios (solar and twice solar composition). The modelling of the planetary atmosphere is a necessary step to know what are the limits of HORuS.  We plot these synthetic models in Fig.~\ref{fig:model_syn_spec}. Further inspection of our synthetic modelling reveals that the depths of the modelled sodium lines are $\approx$~3~$\times$~10$^{-5}$ with respect to the continuum. Therefore, we need to improve the noise level at least by a factor of 30 to be able to confidently detect any \ion{Na}{i} in  55~Cnc~e. Regarding H$\alpha$, our models are consistent with a non-detection of this feature at the resolution of HORuS independently of the  C/O ratio (Fig.~\ref{fig:model_syn_spec}). Thus, ruling out atmospheres with lower mean molecular weight than those of our models requires a better signal-to-noise ratio. In consequence, future observations with an improved signal-to-noise ratio are required in order to be able to confidently detect (3$\sigma$ detection) the atmosphere of 55~Cnc~e.	

\section{Conclusions}

\label{sect_con}

In this work, we have analysed the transit of the super-Earth 55~Cnc~e in the \ion{Na}{i} doublet region to explore the capabilities of HORuS as a high-resolution {\it echelle} spectrograph. The main conclusions of this work can be summarized as follows:

\begin{enumerate}
	\item We estimated uncertainties in the derived radial velocities at about 0.2~km~s$^{-1}$. This is typical for instruments like HORuS, which have not been stabilized thermally, optically, or mechanically.
        \item We confirm that the high sampling of the signal provided by HORuS, with order widths of about 60 pixels in the spatial direction and 6 pixels in the spectral direction, can provide a signal-to-noise ratio of about 1900 per resolution element in a single 450-s exposures for 55 Cnc.
	
	\item Our methodology has allowed us to place a limit on the detection of the planetary signal in the by  \ion{Na}{i} resonance doublet  of ($3.4$~$\pm$~$0.4$)~$\times$~$10^{-4}$, improving by a factor of approximately 10 a previous limit given by \citet{ridh16}.

	\item We have also studied the noise level for the H$\alpha$ line, proving an limit to the  presence of H$\alpha$: ($7$~$\pm$~$1$)~$\times$~$10^{-4}$. 
	\item Our modelling indicated that the signal to noise of our observations has to be increased by a factor of 30 to detect the atmosphere of 55~Cnc~e at the 3$\sigma$ level.  
	\item We have demonstrated that we are able to reach the photon-noise limit of the exoplanetary data using HORuS to a degree that we can detect planets with larger atmospheres (i.e., giant planets). These planets will be easily detected, as indicated by the results of this study.
\end{enumerate}

\section*{Acknowledgements}
We would to thank the anonymous referee for the insightful comments and suggestions that improved the manuscript of the paper. This work was supported by Funda\c{c}\~{a}o para a Ci\^{e}ncia e a Tecnologia (FCT) through the research grants UID/FIS/04434/2019, UIDB/04434/2020 and UIDP/04434/2020. HMT and MRZO acknowledge financial support from the Spanish Ministerio de Ciencia, Innovaci\'{o}n y Universidades through projects AYA2016-79425-C3-2. HMT also acknowledges the FCT - Funda\c{c}\~{a}o para a Ci\^{e}ncia e a Tecnologia through national funds (PTDC/FIS-AST/28953/2017) and by FEDER - Fundo Europeu de Desenvolvimento Regional through COMPETE2020 - Programa Operacional Competitividade e Internacionaliza\c{c}\~{a}o (POCI-01-0145-FEDER-028953). JIGH acknowledges financial support from the Spanish Ministry of Science, Innovation and Universities (MICIU) under the 2013 Ram{\'o}n y Cajal program RYC-2013-14875. CAP, JIGH, and RR acknowledge financial support from the Spanish Ministry project MICIU AYA2017-86389-P. CdB acknowledges the funding of his sabbatical position through the Mexican national council for science and technology (CONACYT grant CVU No. 448248). CdB is also thankful for the support from the Jesus Serra Foundation Guest Program.
%%%%%%%%%%%%%%%%%%%%%%%%%%%%%%%%%%%%%%%%%%%%%%%%%%


\section*{Data Availability}
The data underlying this article will be shared on reasonable request to the corresponding author.

\bibliographystyle{mnras}
\bibliography{55Cnc} % if your bibtex file is called example.bib

%%%%%%%%%%%%%%%%%%%%%%%%%%%%%%%%%%%%%%%%%%%%%%%%%%

%%%%%%%%%%%%%%%%% APPENDICES %%%%%%%%%%%%%%%%%%%%%

\appendix

\section{Extra material}

   \begin{table*}
      %\centering
      \caption{Observational time series for 55~Cnc. We provide Modified Julian Dates (MJD), radial velocities (RVs) for the \ion{Na}{i} and H$\alpha$ wavelengths as well as the effective resolution (in terms of $FWHM$) and the width of the convolution kernel ($FWHM_{\rm k}$) for each individual exposure.} 
      \label{tab:obs_rv_table}
      \begin{tabular}{l c c c c c c c c c}
      \hline
      \hline                                                                                                                                                                                                                         
      mjd  &  t$_{\rm exp}$ &  RV (\ion{Na}{i}) &  $FWHM$~(\ion{Na}{i}) & $FWHM_{\rm k}$ (\ion{Na}{i}) & RV~(H$\alpha$)  & $FWHM$~(H$\alpha$) & $FWHM_{\rm k}$~(H$\alpha$) \\                                                                                                                                                    
        & (s)            & (km~s$^{-1}$)   &   (km~s$^{-1}$)   & (km~s$^{-1}$)            & (km~s$^{-1}$)  & (km~s$^{-1}$) &  (km~s$^{-1}$) \\          
      \hline                                                                                                                                                                                                                         
   58465.01122 & 900 & 27.79 $\pm$ 0.25 & 13.46~$\pm$~0.14 & 5.12 &  27.72 $\pm$ 0.21 & 14.93~$\pm$~0.15 & 4.43\\                                                                                                                                                                                    
   58465.02281 & 900 & 27.53 $\pm$ 0.24 & 13.38~$\pm$~0.13 & 5.32 &  27.57 $\pm$ 0.20 & 14.75~$\pm$~0.15 & 5.02\\                                                                                                                                                                                    
   58465.03939 & 900 & 27.40 $\pm$ 0.23 & 13.14~$\pm$~0.13 & 5.89 &  27.50 $\pm$ 0.19 & 14.50~$\pm$~0.15 & 5.7\\                                                                                                                                                                                    
   58465.05083 & 900 & 27.22 $\pm$ 0.23 & 13.02~$\pm$~0.13 & 6.15 &  27.43 $\pm$ 0.18 & 14.37~$\pm$~0.14 & 6.02\\                                                                                                                                                                                    
   58465.06241 & 450 & 27.28 $\pm$ 0.23 & 12.90~$\pm$~0.13 & 6.40 &  27.51 $\pm$ 0.18 & 14.23~$\pm$~0.14 & 6.34\\                                                                                                                                                                                    
   58465.06864 & 450 & 27.29 $\pm$ 0.23 & 12.83~$\pm$~0.13 & 6.54 &  27.44 $\pm$ 0.18 & 14.16~$\pm$~0.14 & 6.50\\                                                                                                                                                                                    
   58465.07487 & 450 & 27.41 $\pm$ 0.23 & 12.76~$\pm$~0.13 & 6.67 &  27.51 $\pm$ 0.18 & 14.09~$\pm$~0.14 & 6.65\\                                                                                                                                                                                    
   58465.08109 & 450 & 27.60 $\pm$ 0.23 & 12.70~$\pm$~0.13 & 6.79 &  27.71 $\pm$ 0.18 & 14.01~$\pm$~0.14 & 6.82\\                                                                                                                                                                                    
   58465.08732 & 450 & 27.72 $\pm$ 0.22 & 12.63~$\pm$~0.13 & 6.92 &  27.85 $\pm$ 0.18 & 13.94~$\pm$~0.14 & 6.96 \\                                                                                                                                                                                    
   58465.09354 & 450 & 27.91 $\pm$ 0.24 & 12.57~$\pm$~0.13 & 7.02 &  27.73 $\pm$ 0.19 & 13.86~$\pm$~0.14 & 7.16 \\                                                                                                                                                                                    
   58465.09975 & 450 & 27.93 $\pm$ 0.22 & 12.50~$\pm$~0.13 & 7.15 &  28.00 $\pm$ 0.17 & 13.80~$\pm$~0.14 & 7.23 \\                                                                                                                                                                                    
   58465.10598 & 450 & 28.01 $\pm$ 0.22 & 12.43~$\pm$~0.12 & 7.27 &  28.16 $\pm$ 0.17 & 13.79~$\pm$~0.14 & 7.25 \\                                                                                                                                                                                    
   58465.11480 & 450 & 28.10 $\pm$ 0.21 & 12.45~$\pm$~0.12 & 7.23 &  28.23 $\pm$ 0.17 & 13.71~$\pm$~0.14 & 7.40 \\                                                                                                                                                                                    
   58465.12101 & 450 & 28.14 $\pm$ 0.22 & 12.69~$\pm$~0.13 & 6.81 &  28.16 $\pm$ 0.17 & 13.96~$\pm$~0.14 & 6.91 \\                                                                                                                                                                                    
   58465.12724 & 450 & 28.20 $\pm$ 0.22 & 12.94~$\pm$~0.13 & 6.32 &  28.27 $\pm$ 0.17 & 14.19~$\pm$~0.14 & 6.43 \\                                                                                                                                                                                    
   58465.13347 & 450 & 28.14 $\pm$ 0.23 & 13.19~$\pm$~0.13 & 5.78 &  28.17 $\pm$ 0.18 & 14.43~$\pm$~0.14 & 5.87 \\                                                                                                                                                                                    
   58465.13969 & 450 & 28.05 $\pm$ 0.23 & 13.43~$\pm$~0.13 & 5.20 &  28.10 $\pm$ 0.18 & 14.66~$\pm$~0.15 & 5.27 \\                                                                                                                                                                                    
   58465.14591 & 450 & 28.07 $\pm$ 0.24 & 13.68~$\pm$~0.14 & 4.50 &  28.05 $\pm$ 0.19 & 14.90~$\pm$~0.15 & 4.55 \\                                                                                                                                                                                    
   58465.15214 & 450 & 28.13 $\pm$ 0.25 & 13.93~$\pm$~0.14 & 3.65 &  28.13 $\pm $0.19 & 15.13~$\pm$~0.15 & 3.72 \\                                                                                                                                                                                    
   58465.15836 & 450 & 28.25 $\pm$ 0.25 & 14.17~$\pm$~0.14 & 2.56 &  28.19 $\pm$ 0.20 & 15.37~$\pm$~0.15 & 2.55 \\                                                                                                                                                                                    
   58465.16730 & 450 & 28.44 $\pm$ 0.26 & 14.40~$\pm$~0.14 & 0.00 &  28.41 $\pm$ 0.20 & 15.58~$\pm$~0.16 & 0.00 \\                                                                                                                                                                                    
   58465.17353 & 450 & 28.43 $\pm$ 0.27 & 14.30~$\pm$~0.14 & 1.69 &  28.36 $\pm$ 0.20 & 15.48~$\pm$~0.16 & 1.76 \\                                                                                                                                                                                    
   58465.17975 & 450 & 28.30 $\pm$ 0.27 & 14.20~$\pm$~0.14 & 2.39 &  28.24 $\pm$ 0.21 & 15.38~$\pm$~0.15 & 2.49 \\                                                                                                                                                                                    
   58465.18598 & 450 & 28.15 $\pm$ 0.27 & 14.10~$\pm$~0.14 & 2.92 &  28.00 $\pm$ 0.21 & 15.28~$\pm$~0.15 & 3.04 \\                                                                                                                                                                                    
      \hline                                                                                                                                                                                                                         
      \end{tabular}                                                                                                                                                                                                                  
      \end{table*}

\begin{figure}                                                                                                                                                                                         
\includegraphics[width=\columnwidth]{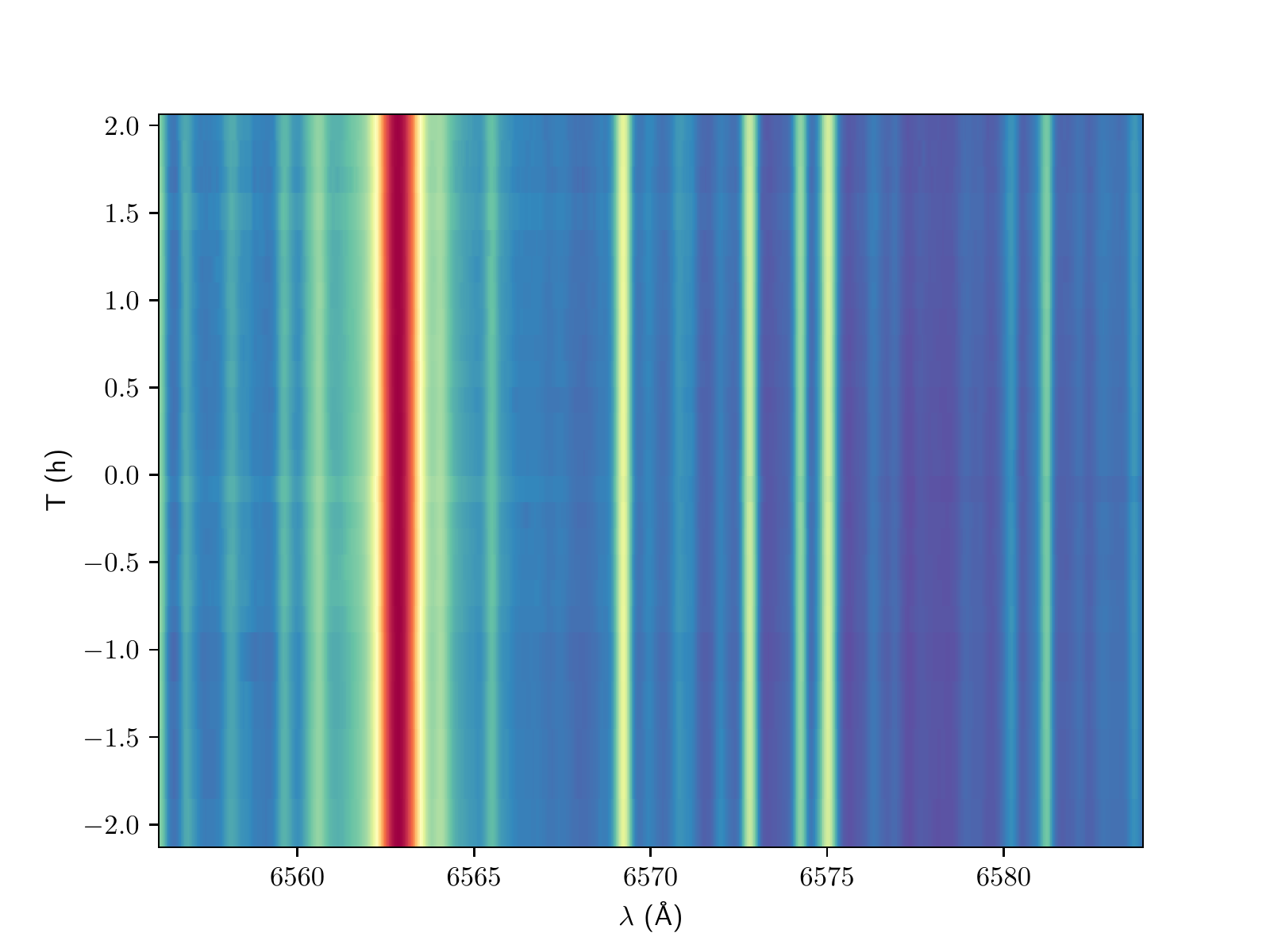}
\caption{Same as Fig.~\ref{fig:scale} but for H$\alpha$}                                                                            
\label{fig:scale_Ha}                                                                                                                                                                    
\end{figure}  

\begin{figure}
  \includegraphics[width=\columnwidth]{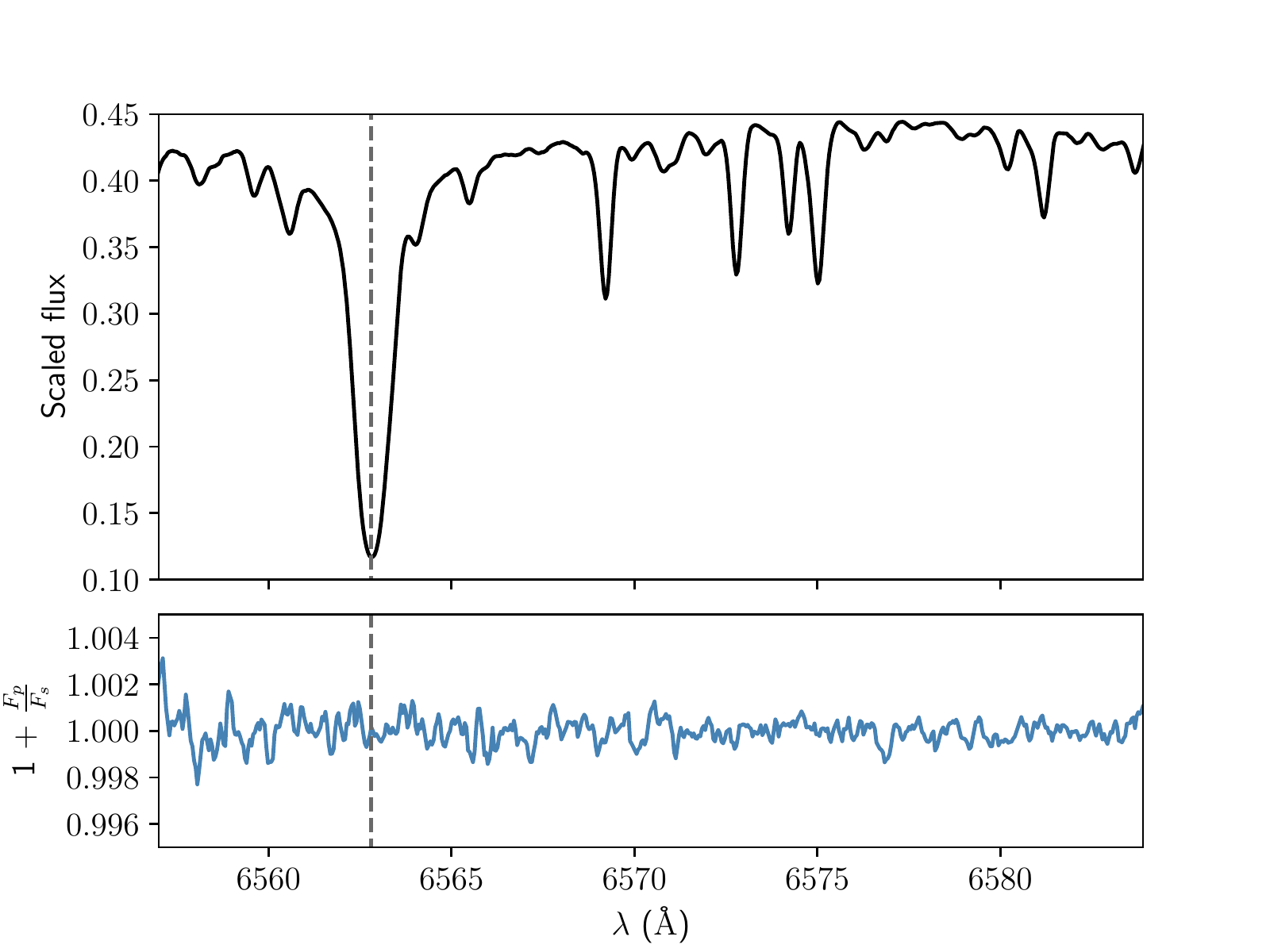}
  \caption{Final master stellar spectrum (top) and final transmission spectrum (bottom) for 55~Cnc~e in the H$\alpha$ region. The H$\alpha$ wavelength is denoted by the vertical dashed line.}
  \label{fig:final_spec_Ha}
  \end{figure}

    \begin{figure}
    \includegraphics[width=\columnwidth]{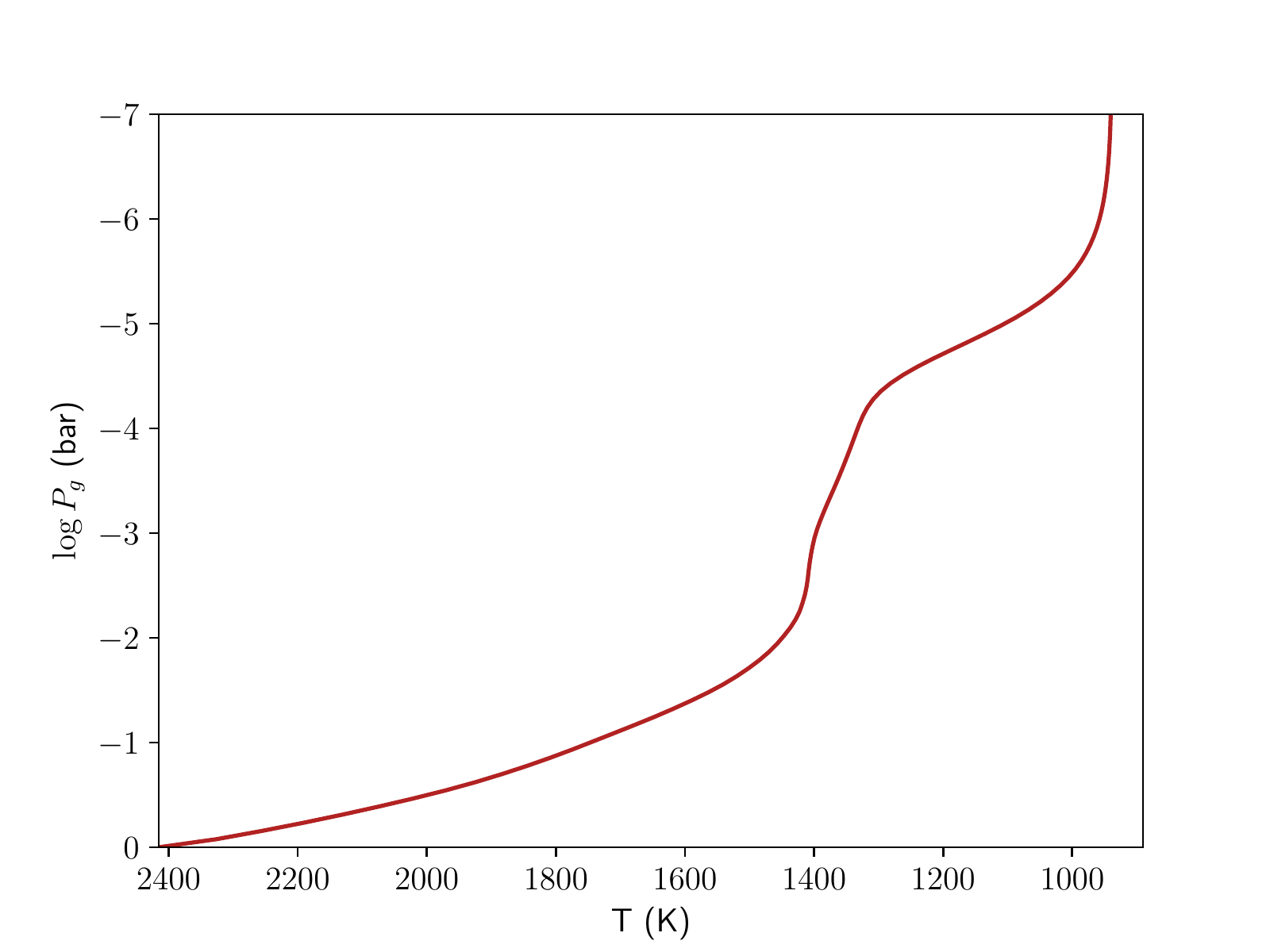}
    \caption{55~Cnc~e atmospheric P-T structure generated with the HELIOS code \citep{mal17,mal19}.}
    \label{fig:model_atm_pt}
    \end{figure}

      \begin{figure}

      \includegraphics[width=\columnwidth]{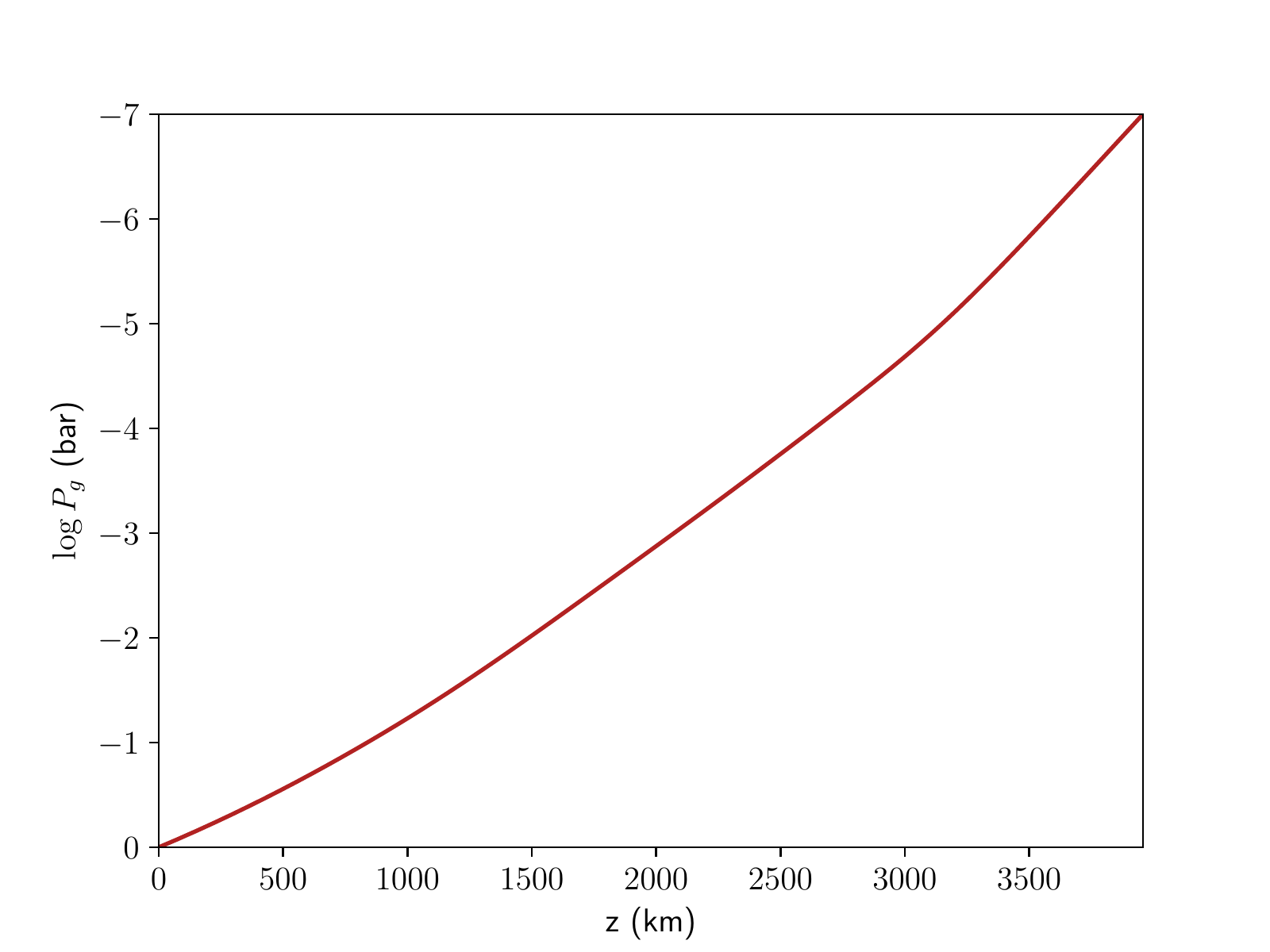}                                                                                                                                            
      \caption{Same as Fig.~\ref{fig:model_atm_pt} but for gaseous pressure as function of altitude (z) } 
      \label{fig:model_atm_pz}                                                                                                                                                                              
      \end{figure}

        \begin{figure}                                                                                                                                                                                          
   \includegraphics[width=\columnwidth]{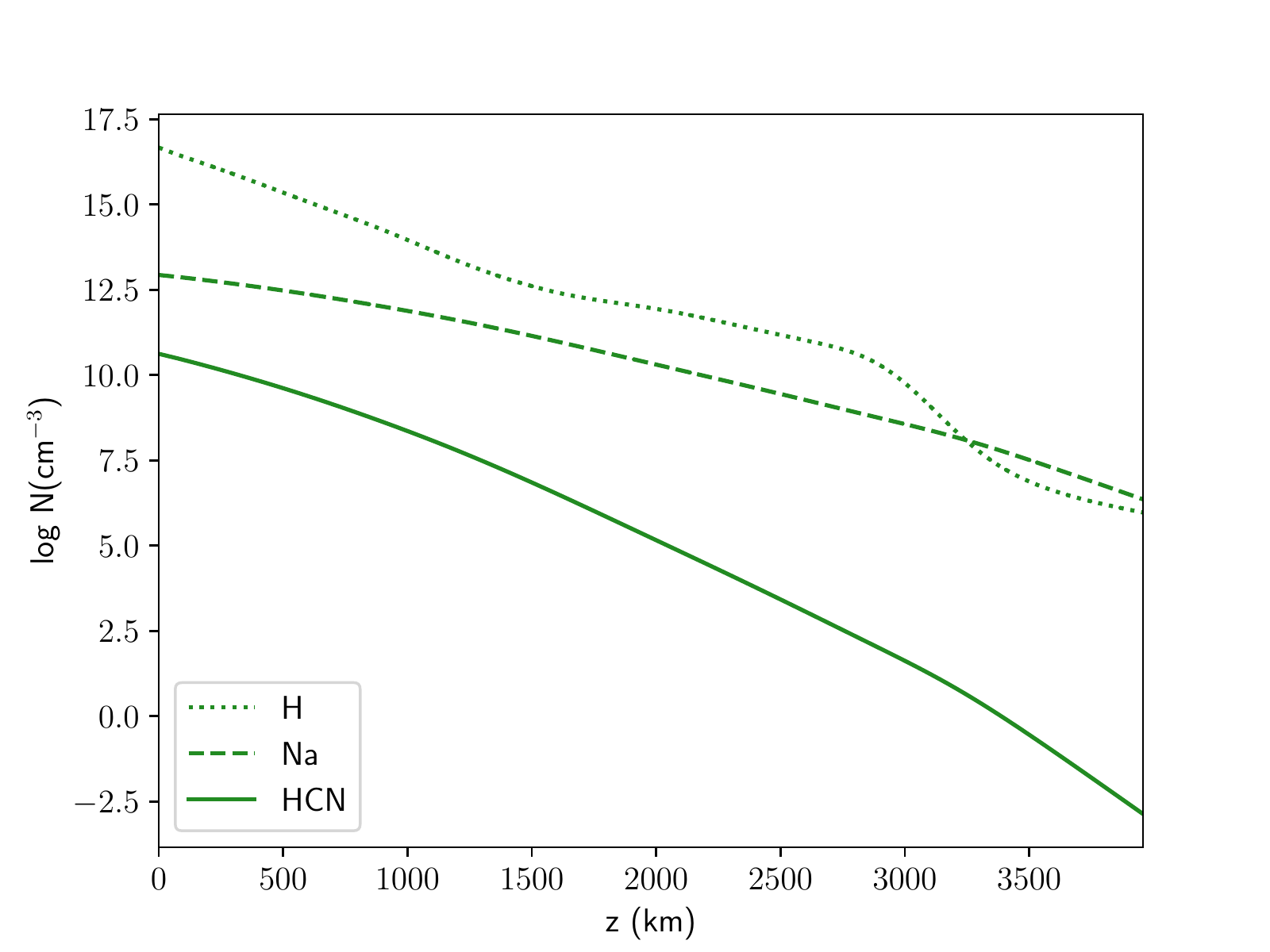}                                                                                                                                                 
    \caption{Number densities of H, HCN, and Na  calculated with Fastchem \citep{sto18} vs the model altitude ($z$). These number densities correspond to a model with a solar C/O ratio.}                                                                                                      
    \label{fig:model_atm_chem}                                                                                                                                                                               
    
\end{figure}
\begin{figure}                                                                                                                                                                                    
  
      \includegraphics[width=\columnwidth]{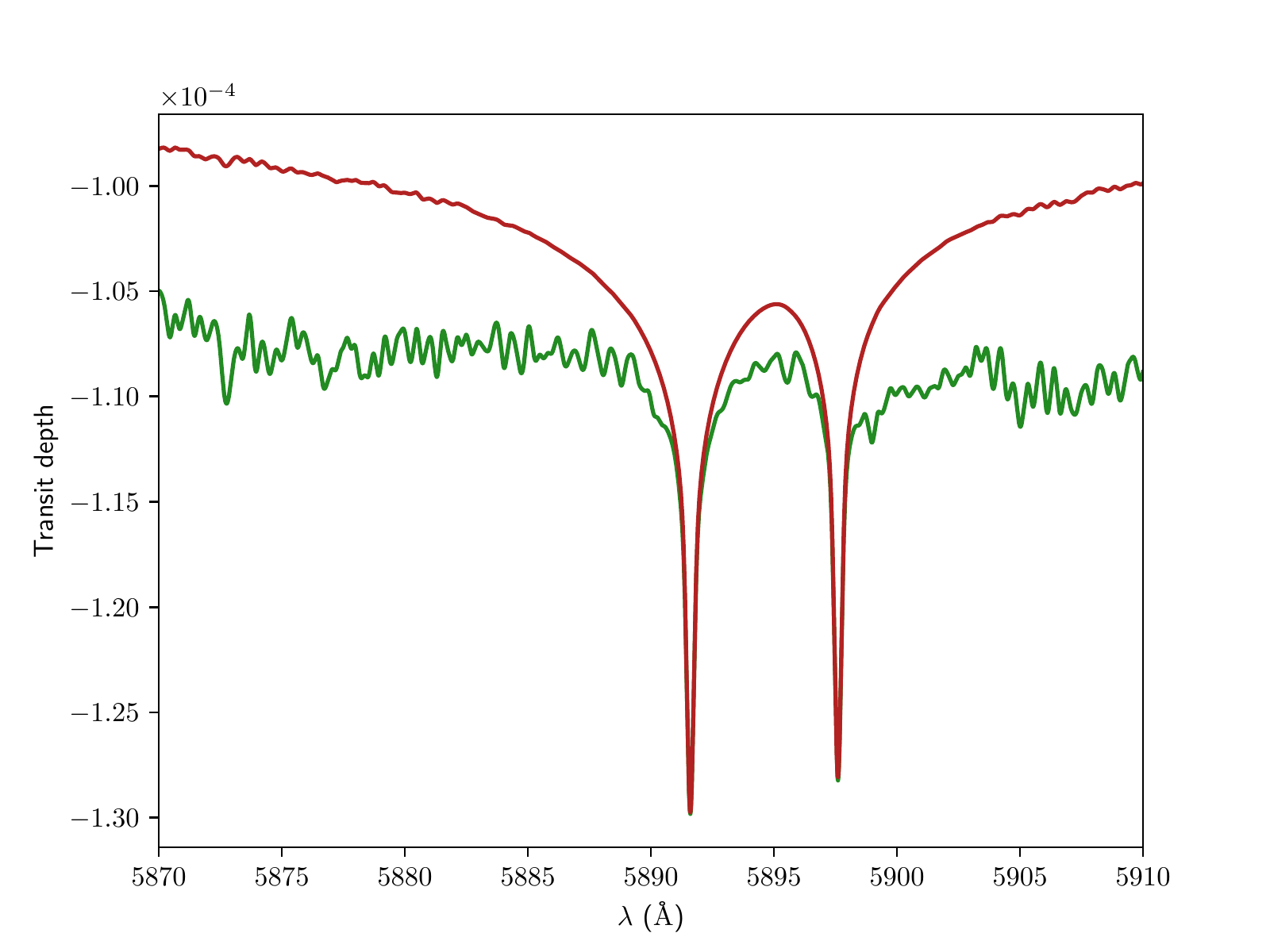}
      \includegraphics[width=\columnwidth]{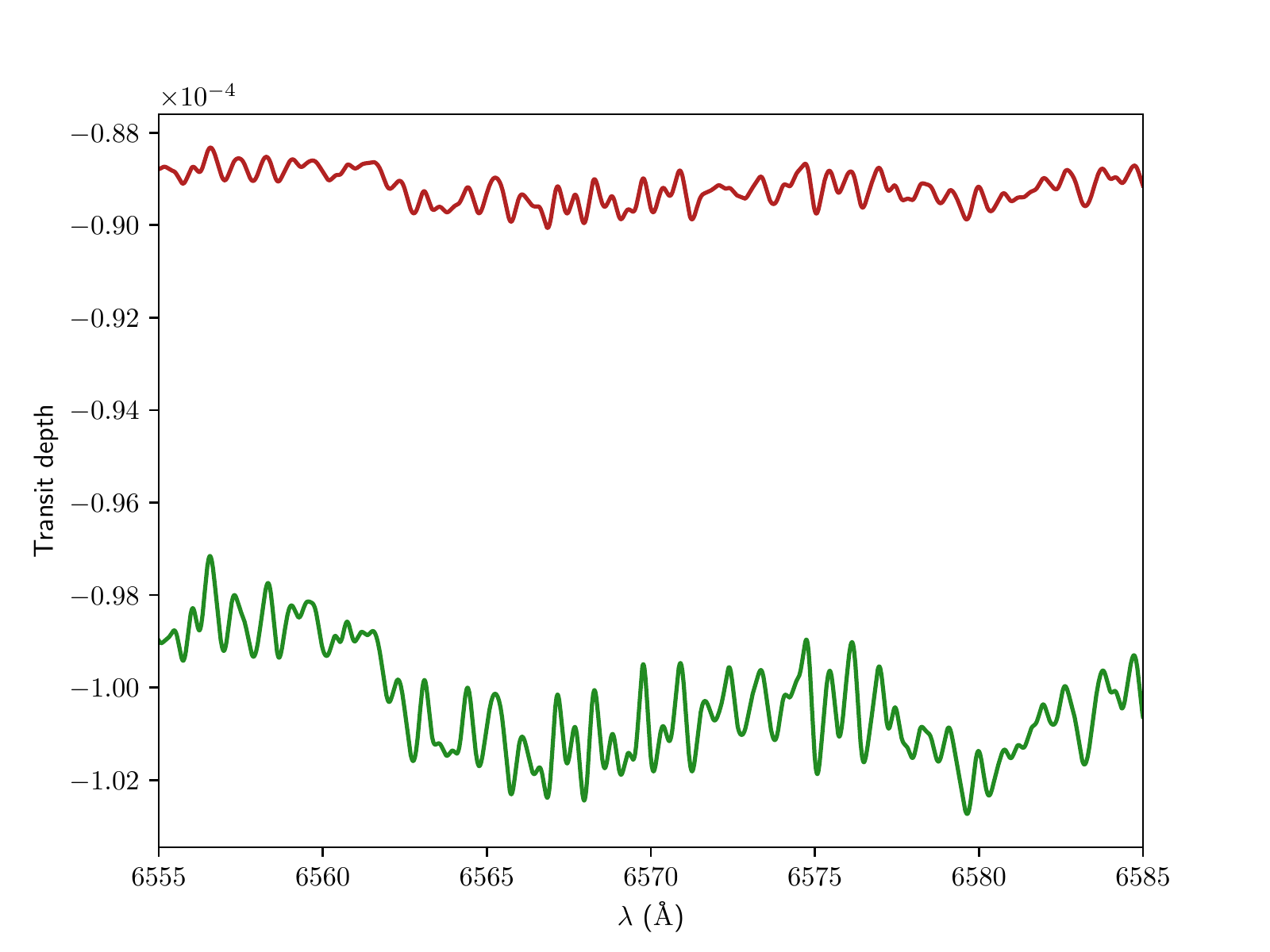} 
      \caption{ Our modelled synthetic spectra around \ion{Na}{i} (top) and H$\alpha$ (bottom) using petitRADTRANS for the optical range.  The green line corresponds to a  model with a solar C/O ratio , whereas the red line represents twice the solar ratio. Both synthetic models have been degraded to the HORuS resolution as provided in Table~\ref{tab:obs_rv_table} }
  
      \label{fig:model_syn_spec}     
    \end{figure}

%%%%%%%%%%%%%%%%%%%%%%%%%%%%%%%%%%%%%%%%%%%%%%%%%%

% Don't change these lines
\bsp	% typesetting comment
\label{lastpage}
\end{document}